\documentclass[aps,prl,twocolumn,superscriptaddress,10pt]{revtex4-1}
\usepackage{chemfig}
\usepackage{amsmath}
\usepackage{amssymb}
\usepackage{color}
\usepackage{times}
\usepackage{epsf}
\usepackage{graphicx}
\usepackage{epsfig}
\usepackage{epstopdf}
\usepackage{dcolumn}
\usepackage{bm}
\usepackage{float}
\usepackage{url}
\usepackage{CJK}

\def\<{\langle}
\def\>{\rangle}
\def\({\left(}
\def\){\right)}
\def\[{\left[}
\def\]{\right]}

\def\cos{\mathop{\mathrm{cos}}\nolimits}

\begin{document}

\title{A Thermal Resistance Network Model for Heat Conduction of Amorphous Polymers}
\author{Jun Zhou}
\affiliation{Center for Phononics and Thermal Energy Science, China-EU Joint Lab for Nanophononics, 
School of Physics Science and Engineering, Tongji University,
Shanghai 200092, China}

\author{Qing Xi}
\affiliation{Center for Phononics and Thermal Energy Science, China-EU Joint Lab for Nanophononics,
School of Physics Science and Engineering, Tongji University, 
Shanghai 200092, China}

\author{Jixiong He}
\affiliation{Department of Mechanical and Aerospace Engineering, North Carolina State University, Raleigh, NC 27695, USA}

\author{Xiangfan Xu}
\affiliation{Center for Phononics and Thermal Energy Science, China-EU Joint Lab for Nanophononics, 
School of Physics Science and Engineering, Tongji University, 
Shanghai 200092, China}

\author{Tsuneyoshi Nakayama}
\affiliation{Center for Phononics and Thermal Energy Science, China-EU Joint Lab for Nanophononics, 
School of Physics Science and Engineering, Tongji University, 
Shanghai 200092, China}
\affiliation{Hokkaido University, Sapporo, Hokkaido 060-0826, Japan}

\author{Yuanyuan Wang}
\email{wangyuanyuan@sspu.edu.cn}
\affiliation{School of Environmental and Materials Engineering, Shanghai Polytechnic University, Shanghai 201209, China}

\author{Jun Liu}
\email{jliu38@ncsu.edu}
\affiliation{Department of Mechanical and Aerospace Engineering, North Carolina State University, Raleigh, NC 27695, USA}

\date{\today}

\begin{abstract}
Thermal conductivities (TCs) of the vast majority of amorphous polymers are in a very narrow range, 0.1 $\sim$ 0.5 Wm$^{-1}$K$^{-1}$, although single polymer chains possess TC of orders-of-magnitude higher. Entanglement of polymer chains plays an important role in determining the TC of bulk polymers. We propose a thermal resistance network (TRN) model for TC in amorphous polymers taking into account the entanglement of molecular chains. 
Our model explains well the physical origin of universally low TC observed in amorphous polymers. The empirical formulae of pressure and temperature dependence of TC can be successfully reproduced from our model not only in solid polymers but also in polymer melts. We further quantitatively explain the anisotropic TC in oriented polymers.

\end{abstract}

\maketitle
Polymers are ubiquitous in a wide range of applications from structure materials to electronics due to their diverse functionality, light weight, low cost, and chemical stability. The low thermal conductivity (TC) of polymers is one of the major technological barrier for the reliability and performance of polymer-based electronics due to the limited heat spreading capability. Significantly different from inorganic materials, the low TC of amorphous polymers is universally confined in a very narrow range, 0.1$\sim$0.5 Wm$^{-1}$K$^{-1}$ \cite{Xu2018}. This feature indicates the possible existence of a universal thermal transport mechanism in amorphous polymers regardless of their distinct chemical structures \cite{Choy1977}. Cahill {\sl et al.} and Xie {\sl et al.} have developed and tested the minimum thermal conductivity model for amorphous polymers, where sound velocity and atomic density govern the TC \cite{Xie2017}. The pressure dependence of TC of  poly(methyl methacrylate) (PMMA) measured by Hsieh {\sl et al.} also agrees with the minimum thermal conductivity model \cite{Hsieh2011}.
Kommandur {\sl et al.} have developed an empirical model to predict temperature-dependent TC of amorphous polymers, where density, monomer molecular weight, and sound velocity govern the dependence \cite{Kommandur2017}. However, these models use bulk properties as inputs, which lack the intrinsic molecular chain details and thus are not able to describe the dependence of TC on temperature, pressure, and orientation simultaneously. 

Amorphous polymer is a three-dimensional (3D) van der Waals (vdW) solid which is a network formed by long one-dimensional (1D) molecular chains \cite{Zallen1998}. A few molecular dynamics (MD) simulations have suggested that a single molecular chain may have a very high TC that is orders-of-magnitude higher than their amorphous counterpart \cite{Henry2008, Liu2012}. This difference is attributed to the fundamental distinction between 3D network and 1D chain. A theoretical model for TC of amorphous polymer which takes into account the structure of 3D network is highly demanded. Both intra-chain and inter-chain thermal transport should be considered in this model, where the intra-chain thermal transport through covalent bond is more efficient than the inter-chain thermal transport via vdW interactions and/or hydrogen bonds.

In this Letter, we propose a thermal resistance network (TRN) model for TC of amorphous polymers, taking into account the interplay of inter-chain and intra-chain thermal transport. Our model successfully describes the value of TC and their relations with chemical structures in various amorphous polymers. Widely used empirical temperature and pressure dependence of TC can be successfully reproduced from our model not only in solid polymers but also in polymer melts. Furthermore, this model is valid to explain the anisotropic TC in oriented polymers nanofibers.   

\begin{figure}[htb]
\includegraphics[width=0.8\linewidth]{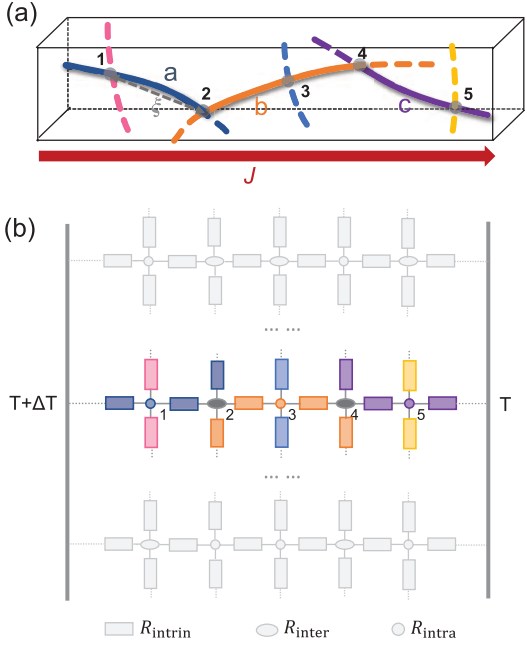}
\caption{(Color online) (a) Illustration of representative unit of an amorphous polymer. Molecular chains are entangled with each other forming a network structure. Possible heat flow trajectory is marked by solid lines where different chains are marked by different colors. Entanglement points are labelled from 1 to 5 in the trajectory. (b) Topologically equivalent 2D TRN corresponding to the trajectory from 1 to 5 shown in (a). Intrinsic resistance of segments, intra-chain resistance at entangled points, and inter-chain resistance are represented by rectangles, circles, and ellipses, respectively.}
\label{fig1}
\end{figure}

Figure\,\ref{fig1} (a) shows a representative unit box with entangled molecular chains that form a random isotropic network. We consider a heat current $J$ flows along the direction of temperature gradient.
Entanglement points between molecular chains are also illustrated. Following the heat current across such a network, we find that the overall TRN consists of three basic elementary resistors: 1) $R_{\rm intrin}$ is the average value of thermal resistance when heat flows through a chain segment between two adjacent points; 2) $R_{\rm intra}$ is the average thermal resistance when heat flows across a point and maintains in the same chain, i.e., intra-chain resistance due to entanglement; and 3) $R_{\rm inter}$ is the average value of interfacial thermal resistance (ITR) when heat flows across a point from one chain to another chain, i.e., the inter-chain resistance.
A typical trajectory of heat current is shown in Fig.\,\ref{fig1}(a) by solid lines. Heat flows from point 1 to point 2, … until point N. Segment (1,2) belongs to chain $a$, segments (2,3) and (3,4) belong to chain $b$, and segment (4,5) belongs to chain $c$. Therefore, heat current flows from chain $a$ to chain $b$ via point 2 and from chain $b$ to chain $c$ via point 4. A topologically equivalent TRN, which is two-dimensional (2D), is shown in Fig.\,\ref{fig1}(b). In an isotropic network, the overall thermal resistance $R$ along the trajectory can be obtained by summing all resistance: 1) overall intrinsic thermal resistance $N\times R_{\rm intrin}$ which are shown as rectangles in Fig.\,\ref{fig1}(b); 2) inter-chain resistance $N_{\rm inter}\times R_{\rm inter}$, where $N_{\rm inter }$ is the average number of inter-chain hopping, which are shown as ellipses; 3) intra-chain resistance $(N-N_{\rm inter})\times R_{\rm intra}$ which are shown as circles. Then we have
\begin{equation}
R=N_{\rm inter}R_{\rm inter}+(N-N_{\rm inter})R_{\rm intra}+NR_{\rm intrin},
\label{equ1}
\end{equation}
where $R_{\rm intrin}=\frac{\xi}{S\kappa_0}$. $\xi$ is the mean distance between two adjacent points, $\kappa_0$ is the intrinsic TC of molecular chains, and $S$ is the cross section of molecular chain. Taking polyethylene (PE) as an example, $S$ is chosen to be 18 $\text{\AA}^2$ \cite{Henry2008} and the simulated $\kappa_0$ is about $10\sim$100 Wm$^{-1}$K$^{-1}$ \cite{Henry2008,Jiang2012,Liu2012}. 
Then the TC of a $d$-dimensional polymer system ($d=2, 3$) can be written by:
\begin{eqnarray}
\kappa &=& \frac{1}{L^{d-2}}\frac{N^{d-1}}{R} \label{equ22}\\
&=& \frac{1}{\left(\xi\overline{\cos\theta}\right)^{d-2}\left[\gamma R_{\rm inter}+(1-\gamma)R_{\rm intra}+R_{\rm intrin}\right]},
\nonumber
\end{eqnarray}
where the size of the system $L=\xi\sum_{i=1}^{N}\cos\theta_{i,i+1}\approx N\xi\overline{\cos\theta}$. $\theta_{i,i+1}$ is the angle between axis and segment (i, i+1), and $\overline{\cos\theta}$ is its average value. 
$\gamma=N_{\rm inter}/N$ is the probability of inter-chain hopping, where $0<\gamma<1$. $\xi$ could be calculated as
\begin{equation}
\xi=\left[\frac{2M_0}{a_0(\overline{\cos\theta})^d\rho}\right]^{\frac{1}{d-1}}.
\label{equ4}
\end{equation}
Here $\rho$ is the mass density, $M_0$ and $a_0$ are the molecular weight and length of the repeating unit, respectively. Eq. (\ref{equ4}) is derived by considering that the total length of trajectory inside a box is $N\xi$, and each segment is entangled with another segment which leads to $\rho=\frac{2N^d\xi M_0}{a_0L^d}$.

\begin{figure}[htp]
\centering
\includegraphics[width=0.9\linewidth]{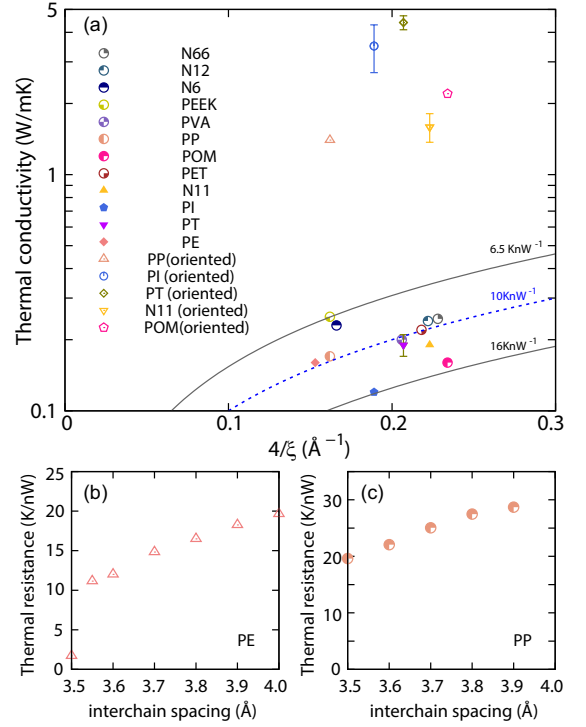}
\caption{(Color online) (a) Calculated TCs versus $4/\xi$ with three different values of $R_{\rm inter}$ are plotted in comparison with the measured TCs of various isotropic polymers and oriented polymers\cite{Zhong2014, Singh2014, Shen2010, Dong2018, Choy1980, Mark2009, David1999, Carvalho1996}. $R_{\rm inter}$  calculated from MD of PE (b) and PP (c) are shown as a function of interchain spacing.}
\label{fig2}
\end{figure}
\begin{table*}
\caption{Structure parameters of typical amorphous polymers and their thermal conductivities. $N_A$ is the Avogadro constant. $\rho$ is from Ref. \cite{Mark2009}, $a_0$ is from Refs.\cite{Okui1987,Mark2009,Tashiro1991}, and $M_0$ is from Refs. \cite{Mark2009}.}
\begin{tabular}{c c c c c c c}
\hline
polymers & $\rho$ & $a_0$ & M$_0\times N_A$ & $\xi$ & $\kappa_{\rm am}$ \\
  &  (gcm$^{-3}$) &  ($\text{\AA}$) &   (gmol$^{-1}$) & ($\text{\AA}$) & (Wm$^{-1}$K$^{-1}$)  \\
\hline
low densidy PE & 0.855 & 1.27 & 28.0  & 26.2 & 0.16 \cite{David1999}  \\
Polyimide (PI)  & 1.42 & 16.0   & 382.0 & 21.1 & 0.12 \cite{Mark2009}\\
polythiophene (PT) & 1.4-1.6  & 7.8  & 194.0 & 21.0 & 0.17-0.21 \cite{Singh2014}   \\
Nylon-11 (N11)  & 1.01 & 15.0 & 183.0 & 17.9 & 0.19 \cite{Mark2009} \\
Poly(methylene oxide) (POM) & 1.42 & 1.93 & 30.0 & 17.1 & 0.16 \cite{David1999} \\
Polypropylene (PP) &  0.85 & 2.17 & 42.1 & 24.6 & 0.17 \cite{David1999} \\
Poly(vinyl alcohol) (PVA) & 1.23-1.33 & 2.52 & 44.0 & 19.4 & 0.2 \cite{Mark2009}  \\
Nylon-6 (N6)  & 0.6-0.7 & 8.6 & 113.2 & 24.1 & 0.23 \cite{Mark2009} \\
Poly(ether ether ketone) (PEEK)  & 1.26 & 10.0 & 288.3 & 24.6 & 0.25 \cite{Mark2009}  \\
Poly(ethylene terephthalate) (PET) &  1.41 & 10.76 & 192.0 & 18.3 & 0.22 \cite{David1999}  \\
Nylon-12 (N12)  & 1.01-1.02 & 16.0  & 198.0 & 18.0 & 0.24 \cite{Mark2009}  \\
Nylon-66 (N66)  & 1.14 & 17.2 & 226.3 & 17.5 & 0.25 \cite{Carvalho1996}  \\
\hline
\end{tabular}
\label{table1}
\end{table*}

\begin{table*}
\caption{Thermophysical properties of typical polymers. $\beta$ of PP is measured at 453 K and others are measured around room temperature.}
\begin{tabular}{c c c c c c c c}
\hline
polymers & $T_g$ & $\alpha_g$ & $\alpha_l$ & $\beta$  & $(1/\kappa_{\rm am})\partial \kappa_{\rm am}/\partial P$ \\
  & (K) &  (10$^{-4}$K$^{-1}$) &  (10$^{-4}$K$^{-1}$) &  (GPa$^{-1}$)  & (GPa$^{-1}$)   \\
\hline
Poly tetra fluoroethylene (PTFE) &  - & - & - & 0.36 \cite{Rae2004} &0.1-0.9 \cite{Ross1984} \\
Nylon-6 (N6) & 320-330 \cite{Mark2009} & - &  3.4-4.0 \cite{handbook2}& - & - \\
Poly(methyl methacrylate)(PMMA) & 387 \cite{handbook2} & 2.7 \cite{handbook2} & 6.1-6.4 \cite{handbook2} & 0.28 \cite{Mark2009} & 0.6-0.7\,\cite{Ross1984}, 0.1-0.2\,\cite{Hsieh2011} \\
Polypropylene (PP), isotactic & 275.5 \cite{Mark2009} & 1.95 \cite{Mark2009} & 4.2\cite{Mark2009} & 1.27 \cite{Mark2009}  & 0.6 \cite{Ross1984}  \\ 
Poly(vinyl acetate) (PVAC) & - & - & - & 0.30 \cite{Mark2009} & 0.9 \cite{Ross1984}  \\
Polystyrene (PS) & 373 \cite{handbook2} & 1.8-2.9 \cite{handbook2} & 4.6-7.2 \cite{handbook2} & 0.27 \cite{Mark2009} & 0.5 \cite{Ross1984} \\
Polycarbonate (PC) & 423 \cite{Mark2009} & 2.6 \cite{Mark2009} & - & 0.26 \cite{Mark2009} & 0.7 \cite{Ross1984}\\
\hline
\end{tabular}
\label{table2}
\end{table*}

Our model is valid for both 2D and 3D polymer systems. Here, we focus our study on 3D amorphous polymers. For isotropic 3D amorphous polymers, $\overline{\cos \theta}=1/2$, thus $\xi=4\sqrt{M_0/(a_0\rho)}$. The probabilities of inter-chain and intra-chain heat transfer at entanglement points are close. Therefore, it is convenient to assume that $\gamma\approx 1/2$. In this case, $R_{\rm intra}$ and $R_{\rm intrin}$ are negligible, as they are much smaller than $R_{\rm inter}$, which is on the order of $10$ K nW$^{-1}$ according to the MD simulations. The reason is that the inter-chain vdW interaction and/or hydrogen bond is much weaker than the covalent bond inside individual chains. Then Eq\,.(\ref{equ22}) becomes
\begin{equation}
\kappa_{\rm am}\approx\frac{4}{\xi R_{\rm inter}}=\sqrt{\frac{\rho a_0}{M_0}}\frac{1}{R_{\rm inter}}.
\label{equ5}
\end{equation}

We evaluate $\xi$ of 12 different polymers (see Table\,\ref{table1}) and plot calculated TCs versus $4/\xi$ in Fig.\,\ref{fig2}(a). We find that $\xi$ ($4/\xi$) lies in a narrow range, 17.1 - 26.2 $\text{\AA}$ (0.15 - 0.23 $\text{\AA}^{-1}$). Then the value of TCs can be explained by choosing $R_{\rm inter}$ being 6.5 $\sim$ 16 KnW$^{-1}$. Especially, TCs of most polymers, except for PE and PI, can be obtained when $R_{\rm inter}\sim$ 10 KnW$^{-1}$. This is because $R_{\rm inter}$ mainly comes from the vdW interactions whose strength should be similar in different polymers.

We further testify $R_{\rm inter}$ of PE and polypropylene (PP) through MD simulations and the results are given in Figs.\,\ref{fig2}(b) and \ref{fig2}(c). The potential between carbons is chosen as $E=\epsilon\left[2\left(\frac{\sigma}{r}\right)^9-2\left(\frac{\sigma}{r}\right)^6\right]$, with $\sigma=4.1 \text{\AA}$ and $\epsilon=2.34$ meV. The polymer models simulated by MD are purely classical systems. Therefore,  we did quantum corrections to the total energy to make sure that the MD simulation temperature is equivalent to a corrected temperature at 300 K \cite{Henry2008}. Furthermore, we simulated the MD temperature-dependent inter-chain resistance and found the dependence is negligible for fixed inter-chain spacing. The rest of the simulation details can be found in Ref. \cite{Liu2012}. The results show that $R_{\rm inter}$ sensitively depends on the inter-chain spacing, which is expected to be below 4.1 $\text{\AA}$ for PE and 4.5 $\text{\AA}$ for PP \cite{handbook2}, as the repulsion between atoms are responsible for the thermal transport between entangled chains below the glass transition temperature ($T_g$). The calculated $R_{\rm inter}$ varies from 2 to 20 KnW$^{-1}$ and from 20 to 30 KnW$^{-1}$ for PE and PP, respectively, when the inter-chain spacing varies from 3.5 to 4 $\text{\AA}$. These values are very close to the values required in Fig. \ref{fig2}(a), considering that the models of polymer chains in MD are oversimplified compared to the real polymers. Therefore, our model is valid and it successfully explains the origin of the small difference of TC of polymers with completely different chemical structures. It should be pointed out that the overlap area between entangled chains are very difficult to determine because of the complicated chemical structures. In our calculations, we assumed that the overlapping area of PE and PP molecular chains are 4$\times$12.7$\text{\AA}^2$ and 4$\times$11 $\text{\AA}^2$, based on the Kuhn length of each polymer, respectively. 

The temperature dependence of TC is derived from Eq. (\ref{equ5}):
\begin{equation}
\frac{1}{\kappa_{\rm am}}\frac{\partial \kappa_{\rm am}}{\partial T}=-\frac{\alpha}{2}-\frac{\partial\text{ln}R_{\rm inter}}{\partial T},
\end{equation}
where $\alpha=-(1/\rho)\partial \rho/\partial T$ is the thermal expansion coefficient. An exact temperature dependence of $R_{\rm inter}$ requires further comprehensive simulations. Here, we assume the temperature dependence of $R_{\rm inter}$ will obey the general trend predicted by the diffuse mismatch model (DMM). In DMM,  $R_{\rm inter}$ gradually decreases with temperature at low temperature and finally saturates near room temperature and this effect is mainly attributed to the temperature-dependent heat capacity \cite{Reddy2005}. Therefore, $-\partial\text{ln}R_{\rm inter}/\partial T$ is positive at low temperature and approaches zero near room temperature. Since $-\alpha/2<0$, the competition between these two terms determines the temperature dependence of TC. There is a discontinuity of $\alpha$ at $T_g$ \cite{Boyer1944} where their values are noted as $\alpha_g$ and $\alpha_l$ below and above $T_g$, respectively, as shown in Table\,\ref{table2}. 
When $T<T_g$, $\alpha_g$ is small and $-\partial\text{ln}R_{\rm inter}/\partial T$ is large.
If we assume $R_{\rm inter}\propto T^{-\delta}$ and neglect $-\alpha_g/2$, then TC gradually increases with temperature as
\begin{equation}
\frac{\kappa_{\rm am}(T)}{\kappa_{\rm am}(T_g)}\approx\left(\frac{T}{T_{g}}\right)^{\delta}, T<T_{g}.
\end{equation} 
This is in consistent with the empirical formula $\frac{\kappa_{\rm am}}{\kappa_{\rm am}(T_g)}=\left(\frac{T}{T_{g}}\right)^{0.22}$ \cite{handbook2} when $\delta=0.22$. When $T>T_g$, $R_{\rm inter}$ is almost independent with temperature, and $-\alpha_l/2$ is dominant which results in a linear decrease of TC as
\begin{equation}
\frac{\kappa_{\rm am}(T)}{\kappa_{\rm am}(T_g)}\approx\left[\left(1+\frac{\alpha_l T_g}{2}\right)-\frac{\alpha_l T_g}{2}\left(\frac{T}{T_g}\right)\right], T>T_{g}.
\label{equ77}
\end{equation} 
Here, $\alpha_l T_g$ is $0.1-0.3$ as shown in Table\,\ref{table2}. This is close to the empirical relation $\frac{\kappa_{\rm am}(T)}{\kappa_{\rm am}(T_g)}=1.2-0.2\frac{T}{T_g}$ \cite{handbook2}. 

The pressure dependence of $\kappa_{\rm am}$ at fix temperature can also be derived from Eq. (\ref{equ5}):
\begin{equation}
\frac{1}{\kappa_{\rm am}}\frac{\partial\kappa_{\rm am}}{\partial P}=\frac{\beta}{2}-\frac{\partial\text{ln}R_{\rm inter}}{\partial P},
\end{equation}
where $\beta=(1/\rho)\partial \rho/\partial P$ is the compressibility whose values are shown in Table \ref{table2} \cite{Ross1984}.  We are not able to calculate $\partial \text{ln}R_{\rm inter}/ \partial P$ at current stage. We speculate that $R_{\rm inter}$ decreases with increasing pressure, due to a stronger entanglement and/or decreased inter-chain distance under pressure. We pointed out that $\frac{1}{\kappa_{\rm am}}\frac{\partial\kappa_{\rm am}}{\partial P}$ is on the order of 0.1-1 GPa$^{-1}$ and is slightly larger than $\beta/2$. This is consistent with the values in Table \ref{table2}.

\begin{figure}[tp]
\centering
\includegraphics[width=0.8\linewidth]{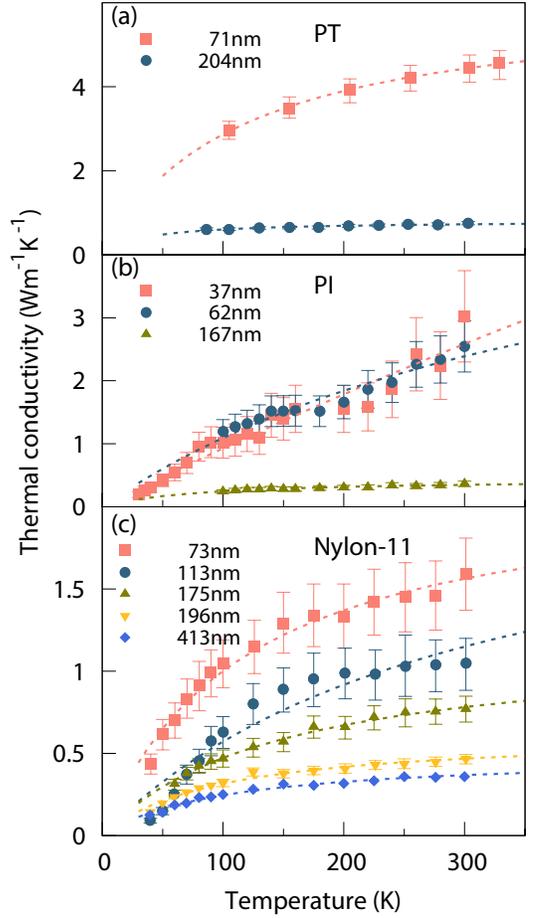}
\caption{(Color online) Temperature dependence of $\kappa_\parallel$ of orientied (a) PT \cite{Singh2014}, (b) PI \cite{Dong2018}, and (c) Nylon-11 \cite{Zhong2014} nanofibers with different diameters.  Dots represent experimental data and dashed lines are fitted by Eq.\,(\ref{equ8}).}
\label{fig3}
\end{figure}

\begin{figure}[tp]
\centering
\includegraphics[width=0.9\linewidth]{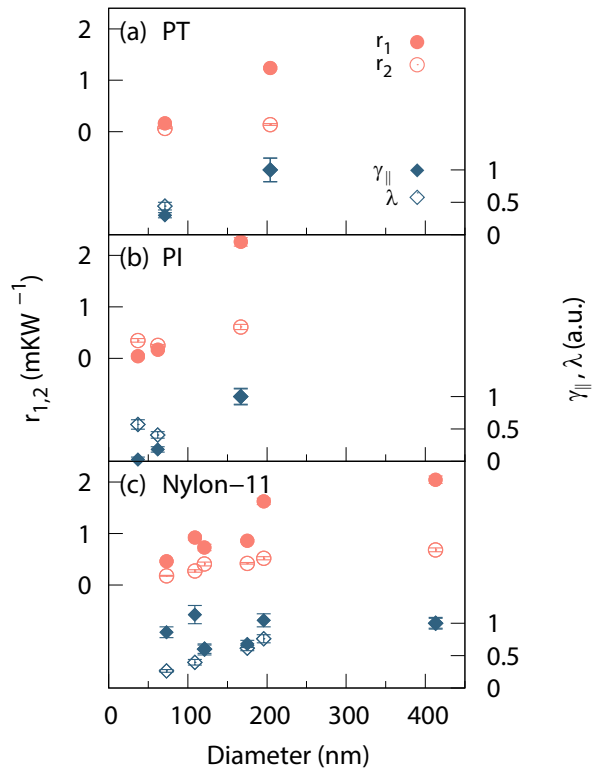}
\caption{(Color online) Fitting parameters $r_{1}$ and $r_{2}$ versus the diameter of polymer nanofibers. $\gamma_{\parallel}$ and $\lambda$ extracted from $r_{1}$ and $r_{2}$ are plotted in arbitrary unit to the right y-axis.}
\label{fig4}
\end{figure}

   
We now study the anisotropic TC of oriented polymers. Many experiments have shown that TC along oriented direction ($\parallel$) is much larger than $\kappa_{\rm am}$ as shown in Fig.\,\ref{fig2}(a) and Table \ref{table1}. TC in perpendicular direction ($\perp$) is smaller \cite{Lu2016}. In this case, $\overline{\cos\theta_{\parallel}}>1/2$ and $\overline{\cos\theta_{\perp}}<1/2$, where $\theta_{\parallel}$ and $\theta_{\perp}$ are the average angles of chain segments with respect to the direction along and perpendicular to the orientation, respectively. The anisotropic inter-chain hopping possibility ($\gamma_\parallel$) is smaller than 1/2. Then we have
\begin{equation}
\kappa_{\parallel}=\frac{\overline{\cos\theta_{\parallel}}}{\xi\overline{\cos\theta_{\perp}}^2\left[\gamma_\parallel R_{\rm inter}+(1-\gamma_\parallel)R_{\rm intra}+R_{\rm intrin}\right]}.
\label{equ6}
\end{equation}
It is clear that the increase of TC comes from the increase of $\overline{\cos\theta_{\parallel}}/\overline{\cos\theta_{\perp}}^2$ and decrease of $\gamma_{\parallel}$. In a highly oriented polymer, $\gamma_{\parallel}\ll 1$ and $\overline{\cos\theta_{\parallel}}\approx 1$, Eq.\,(\ref{equ6}) goes to the limit form as $\kappa_{\parallel}\rightarrow [\xi\overline{\cos\theta_{\perp}}^2 (R_{\rm intra}+R_{\rm intrin})]^{-1}$. It means that the TC of highly oriented polymers is dominated by the intrinsic TC of molecular chains and $R_{\rm inter}$ is negligible. 
It is reasonable to assume that $\kappa_0\approx \chi T$ near room temperature according to MD simulations \cite{Jiang2012} where $\chi$ is a constant. Then Eq. (\ref{equ6}) can be simplified as
\begin{equation}
\kappa_{\parallel}=\frac{1}{r_{1}+\frac{r_{2}}{T/T_0}} ,
\label{equ8}
\end{equation}
where $r_1$ and $r_2$ are two parameters which can be written as $r_{1}=\xi\gamma_{\parallel}\left[\lambda (R_{\rm inter}-R_{\rm intra})+R_{\rm intra}\right]$ and $r_{2}=\xi^2\lambda /(S\chi T_0)$, with $\lambda={\overline{\cos\theta_\perp}^2}/{\overline{\cos\theta_\parallel}}$ and $T_{0}$=300 K.
We use Eq.\,(\ref{equ8}) to fit the experimental measured $\kappa_{\parallel}$ of PT, PI, and Nylon-11 nanofibers with different diameters in Fig. \ref{fig3}. Our formula is in excellent agreement with the experimental data where the fitted $r_1$ and $r_2$ are shown in Fig. \ref{fig4}. It is interesting that $\lambda$ and $\gamma_{\parallel}$ with arbitrary unit can be deduced from $r_1$ and $r_2$. They are also presented in Fig. \ref{fig4} to the right y-axis. For nanofibers with large diameters, $r_{1}$ is significantly larger than $r_{2}$, then the temperature dependence of $\kappa_{\parallel}$ is weak which is similar to the case of isotropic polymers. We find that both $r_{1}$ and $r_{2}$ decrease with decreasing diameter, while $r_{1}$ decreases more rapidly than $r_{2}$. This is because $r_1$ includes both $\lambda$ and $\gamma_{\parallel}$ that decrease with the decreasing of diameter, while $r_2$ does not include $\gamma_{\parallel}$. As a result, $r_{2}$ becomes comparable with $r_{1}$ for diameters below 100nm, then $\kappa_{\parallel}$ shows a stronger temperature dependence. In ultra-thin nanofibers with diameter smaller than 50nm, $r_{2}/(T/T_{0})>>r_1$ is satisfied, one can find that $\kappa_{\parallel}\propto T$.



Finally, we extend our model to discuss other effects on TC without loss of generality:

1) Crystallinity effect: Semi-crystalline polymers are composed of crystalline phase and amorphous phase. Since the crystalline phase is formed by ordered molecular chains, their TC is similar to crystal solids, which has been studied a lot through measurements of polymers with high crystallinity \cite{Wang2013}. The difficulties in predicting TC of semi-crystalline polymers still lies in the poor understanding of amorphous phase \cite{Lu2018}, and our model will serve as an effective approach to evaluate that. 

2) Crosslinking effect: It is known that crosslinking could enhance TC of amorphous polymers \cite{Kikugawa2013, Xiong2017,Rashidi2017,Yamamoto1971}. Under the framework of our model, the crosslink bonds can be seen as altering some entangled points via vdW interaction by linked points via real bonding, which will decrease $R_{\rm inter}$, thus increase TC.

3) Branched effect: Branched polymers are found to possess lower TC than polymers with single linear chains due to a lower density \cite{Fuller1971}. This can be easily understood that a lower density $\rho$ results in a larger $\xi$, thus TC will be reduced.   

In summary, we proposed a thermal resistance network model that describes well the thermal conductivity of amorphous polymers. The entangled network structure and the interplay between intra-chain and inter-chain heat transfer are considered in our model. The fundamental mechanism of a universally low thermal conductivity of polymers are found to be the similar mean distance between entangled points and the similar inter-chain resistance due to vdW interaction. Our model successfully reproduce the empirical temperature dependence and pressure dependence
of thermal conductivity not only in solid polymers but also in polymer metls. Moreover, the experimentally observed anisotropic TC can be quantitatively explained by our model.

Acknowledgments. This work was supported by National Key R{\&}D Program of China (No. 2017YFB0406004), National Natural Science Foundation of China (No. 11890703), and Shanghai Key Laboratory of Special Artificial Microstructure Materials and Technology (2019$\sim$2022). This work was also supported by the Faculty Research and Professional Development Fund at North Carolina State University.


\begin{thebibliography}{99}

\bibitem{Xu2018} X. Xu, {\sl et al.}, Adv. Mater. {\bf 30}, 1705544 (2018).

\bibitem{Choy1977} C. L. Choy, Polymer {\bf 18}, 984 (1977).

\bibitem{Xie2017} X. Xie, {\sl et al.}, Phys. Rev. B {\bf 95}, 035406 (2017).

\bibitem{Hsieh2011} W.-P. Hsieh, {\sl et al.}, Phys. Rev. B {\bf 83}, 174205 (2011).

\bibitem{Kommandur2017} S. Kommandur and S. K. Yee, J. Polym. Sci. Polym. Phys. {\bf 55}, 1160 (2017).

\bibitem{Zallen1998} R. Zallen, The Physics of Amorphous Solids (Wiley, New York, 1998) p. 107-133.

\bibitem{Henry2008} A. Henry and G. Chen, Phys. Rev. Lett. {\bf 101}, 235502 (2008).

\bibitem{Liu2012} J. Liu and R. Yang, Phys. Rev. B {\bf 86}, 104307 (2012).


\bibitem{Zhong2014} Z. Zhong, {\sl et al.}, Nanoscale {\bf 6}, 8283 (2014).

\bibitem{Singh2014} V. Singh et al., Nat. Nanotech. {\bf 9}, 384 (2014).

\bibitem{Shen2010} S. Shen, {\sl et al.}, Nat. Nanotech. {\bf 5}, 251 (2010).



\bibitem{Dong2018} L. Dong, et al., Nat. Sci. Rev. {\bf 5}, 500 (2018).

\bibitem{Choy1980}	C. L. Choy, F. C. Chen, and W. H. Luk, J. Polym. Sci. Polym. Phys. {\bf 18}, 1187 (1980).


\bibitem{Jiang2012} J. W. Jiang, {\sl et al.}, J. Appl. Phys. {\bf 111}, 124304 (2012).

\bibitem{Mark2009} J. E. Mark, Polymer Data Handbook (Oxford University press, Oxford, 2009).

\bibitem{David1999} D. J. David and A. Misra, Relating Materials Properties to Structure with MATPROP Software: Handbook and Software for Polymer Calculations and Materials Properties(CRC Press, Boca Raton,1999) p. 531.

\bibitem{Carvalho1996} G. D. Carvalho, E. Frollini, and W. N. D. Santos, J. Appl. Polym. Sci. {\bf 62}, 2281 (1996).

\bibitem{Rae2004} P. Rae and D. M. Dattelbaum, Polymer {\bf 45}, 7615 (2004).

\bibitem{Okui1987} N. Okui and T. Sakai, Polym. Bull. {\bf 17}, 79 (1987).

\bibitem{Tashiro1991} K. Tashiro, {\sl et al.}, J. Polym. Sci. Polym. Phys. {\bf 29}, 1223 (1991).

\bibitem{Boyer1944} R. F. Boyer and R. S. Spencer, J. Appl. Phys. {\bf 15}, 398 (1944).

\bibitem{Reddy2005} P. Reddy, K. Castelino, and A. Majumdar, Appl. Phys. Lett. {\bf 87}, 211908 (2005).

\bibitem{handbook2} D. W. Van Krevelen and K. Te Nijenhuis, Properties of Polymers, Fourth edition, (Elsevier, Oxford, 2009) p.\,74, 92-94, 647. 


\bibitem{Ross1984}	R. G. Ross, {\sl et al.}, Rep. Prog. Phys. {\bf 47}, 1347 (1984).

\bibitem{Lu2016} Y. Lu, {\sl et al.}, Acs Macro Lett. {\bf 5}, 646 (2016)


\bibitem{Wang2013}	X. Wang, {\sl et al.}, Macromolecules {\bf 46}, 4937 (2013).

\bibitem{Lu2018} T. Lu, {\sl et al.},  J. Appl. Phys. {\bf 123},015107 (2018).

\bibitem{Kikugawa2013}	G. Kikugawa, {\sl et al.}, J. Appl. Phys. {\bf 114}, 034302 (2013).

\bibitem{Xiong2017} X. Xiong, {\sl et al.}, J. Appl. Phys. {\bf 122}, 035104 (2017).

\bibitem{Rashidi2017} V. Rashidi, {\sl et al.}, J. Phys. Chem. B {\bf 121}, 4600 (2017).

\bibitem{Yamamoto1971} O. Yamamoto and H. Kambe, Polym. J. {\bf 2}, 623 (1971).

\bibitem{Fuller1971} T. R. Fuller and A. L. Fricke, J. Appl. Polym. Sci. {\bf 15}, 1729 (1971).

\end{thebibliography}
\end{document}